\documentclass[reprint]{revtex4-1}
\usepackage[english]{babel}
\usepackage{graphicx}
\usepackage[font={small},justification=centerlast]{caption}
\usepackage{amsmath}
\usepackage{float}
\usepackage{wrapfig}
\usepackage{bm}
\usepackage{mathtools}
\usepackage{xr}
\usepackage{amssymb}
\usepackage{color}
\usepackage{epstopdf}

\draft 

\begin{document}


\title{Phase and vacancy behaviour of hard ``slanted'' cubes}



\author{R. van Damme$^1$}
\author{B. van der Meer$^1$}
\author{J. J. van den Broeke$^1$}
\author{F. Smallenburg$^2$}
\author{L. Filion$^1$}
\affiliation{${}^1$ Soft Condensed Matter, Debye Institute for Nanomaterials Science, Utrecht University, Princetonplein 5, 3584 CC Utrecht, The Netherlands\\
${}^2$ Institut f\"ur Theoretische Physik II: Weiche Materie, Heinrich-Heine-Universit\"at D\"usseldorf, Universit\"atsstr. 1, 40225 D\"usseldorf, Germany}

\date{\today}

\begin{abstract}
We use computer simulations to study the phase behaviour for hard, right rhombic prisms as a function of the angle of their rhombic face (the ``slant" angle). More specifically, using a combination of event-driven molecular dynamics simulations, Monte Carlo simulations, and free-energy calculations, we determine and characterize the equilibrium phases formed by these particles for various slant angles and densities. Surprisingly, we find that the equilibrium crystal structure for a large range of slant angles and densities is the simple cubic crystal - despite the fact that the particles do not have cubic symmetry. Moreover, we find that the equilibrium vacancy concentration in this simple cubic phase is extremely high and depends only on the packing fraction, and not the particle shape.  At higher densities, a rhombic crystal appears as the equilibrium phase. We summarize the phase behaviour of this system by drawing a phase diagram in the slant angle - packing fraction plane.

\end{abstract}


\maketitle 

\section{Introduction}
\label{sec:Introduction}

Recent years have seen a number of studies into the phase behaviour of hard, colloidal particles with different shapes. One of the reasons for the interest in these systems can be seen clearly when comparing virtually any two of these papers: shape matters. Depending on the shape of particles, it is possible to form a vast array of different phases (see e.g. Refs. \cite{haji2009disordered,damasceno2012predictive,agarwal2011mesophase,dussi2016entropy,marechal2012frustration,ni2012phase,gantapara2013phase}), and small differences in the shape of the constituent particles can evidently mean large
differences in their collective phase behaviour (see e.g. Refs. \cite{marechal2012frustration,ni2012phase,gantapara2013phase}). As advances in colloidal synthesis provide an ever-increasing control over the shape of the particles, knowledge of the shape-dependent phase behaviour becomes
increasingly valuable.

An interesting example of how shape can lead to novel and unexpected phase behaviour was reported for hard cubes in Ref. \cite{Smallenburg2012}.  Here, it was shown that particle shape can have an extremely intriguing effect on the concentration and realization of defects in colloidal systems.  It is well known that at finite temperatures, all equilibrium solid phases possess point defects such as vacancies and interstitials. These defects arise as the free energy of the solid phase is always minimized by a small but nonzero fraction of such defects. In typical, one-component crystals, the equilibrium fraction of such defects is extremely low - generally on the order of one vacancy for every ten thousand or more particles.  As a result, most studies of phase behaviour can (and do) safely ignore such defects.

A stunning exception to this rule was demonstrated for hard cubes on a simple cubic lattice \cite{Smallenburg2012}. Specifically it was shown that near melting, the simple cubic crystal formed by hard cubes was riddled with vacancies, reaching equilibrium vacancy concentrations of up to 6\% - three orders of magnitude larger than for a typical ionic, metallic, or colloidal crystal. Interestingly, the vacancies manifest as extended point defects: each vacancy is shared over a string of lattice sites along one of the three main crystal directions \cite{Smallenburg2012}.  

\begin{figure}
\includegraphics[width=0.9\linewidth]{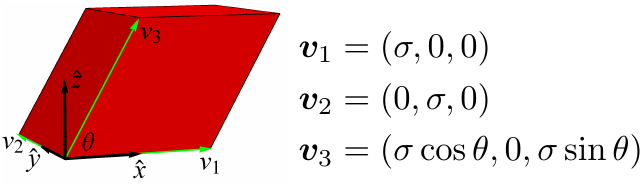}
\caption{
\label{fig:TiltedCubeVectors} A slanted cube with edge length $\sigma$ and slant angle $\theta$.}
\end{figure}

In light of previous works which showed that phase behaviour can be highly sensitive to the exact shape of particles \cite{marechal2012frustration,ni2012phase,gantapara2013phase}, one might reasonably ask how robust these defects are to the particle shape. As a first step in this direction, Gantapara {\it et al.} \cite{gantapara2013phase} examined the effect of truncating the cubes on the phase behaviour of this system. For this system, where the particles retain their cubic symmetry, they found that the defects were largely robust to such a particle deformation.  Here we attack this question from a different angle, and modify the particles in a continuous way that removes their cubic symmetry - and explore how this modification affects both their stable crystalline phases as well as the way in which defects manifest in this system.  

Specifically, we examine the phase behaviour of hard, right rhombic prisms as shown in Fig. \ref{fig:TiltedCubeVectors}, with ``slant'' angles between $66^\circ$ and $90^\circ$.  As these particles are simply formed by shearing a simple cube, we expected a priori that the crystalline phase in this system would  also be a sheared simple cubic crystal phase.  Here we show that this assumption is incorrect, and that the crystal formed near melting is instead a simple cubic phase - again filled with vacancies.  Finally, we will show that the equilibrium vacancy concentration for this sheared cubic system is, amazingly, essentially independent of the degree of slanting of the particles.

\section{Methods}
\label{sec:Methods}
\subsection{Model system} 
We examine the phase behaviour of hard, right rhombic prisms as shown in Fig. \ref{fig:TiltedCubeVectors}.  The shape of these particles is completely described by the edge length $\sigma$, and a single angle $\theta$ which corresponds to the angle of the rhombus.  This angle we refer to as the ``slant angle''. For simplicity, and to emphasize the similarity to cubes, we will refer to these particles as ``slanted cubes''.  Note that a slant angle of $\theta=90^\circ$ corresponds to a cube and forms the upper limit of $\theta$. The particles interact only via hard-core interactions: the interaction potential is zero for all configurations without overlaps and infinity when overlaps are present.   

We focus on slant angles that are likely to result in crystals which can form extended defects, and thus consider only angles between $\theta =66^\circ$  and $\theta =90^\circ$.  This avoids angles close to $\theta = 60^\circ$, which would accommodate a wide range of crystal structures incorporating layers with rhombic tilings,  (such as those described in Ref. \cite{whitelam2012random}).

\subsection{Free-energy calculations}
In order to draw the phase diagram of this system, we use a combination of event-driven molecular dynamics simulations (EDMD), Monte Carlo (MC) simulations and free-energy calculations. In the simulations, we use the Separating Axis Theorem \cite{Gottschalk1996a} to detect overlaps and predict collisions in the systems. For the EDMD simulations, we follow the method described in Refs. \cite{Smallenburg2012,hernandez2007discontinuous}.  In the following we describe the free-energy calculations we use to study the fluid and crystal phases of the slanted cubes.

\subsubsection{Fluid free energy}
In order to calculate the free energy of the fluid phase for a given slant angle $\theta$, we first measure the equation of state of the fluid. Specifically, we perform MC simulations in the $NpT$ ensemble, i.e. at constant number of particles $N$, constant pressure $p$, and constant temperature $T$, and measure the average number density $\rho$ in each simulation. Alternatively, we performed EDMD simulations at constant $N$, volume $V$, and temperature $T$, and measured the pressure. The resulting relation $p(\rho)$ between the pressure  and the density can be used to calculate the Helmholtz free energy $F$ of the fluid via thermodynamic integration, using the ideal gas as a reference state \cite{frenkel2001understanding}. Specifically, 
\begin{equation}
\dfrac{\beta F(\rho)}{N}=\dfrac{\beta F_\mathrm{id}(\rho)}{N}+\beta \int^{\rho }_0 d\rho' \dfrac{p(\rho') -\rho' /\beta }{\rho'^ 2},
\end{equation}
where $\beta = 1/k_B T$ with $k_B$ Boltzmann's constant and $F_\mathrm{id}(\rho)$ the ideal-gas free energy.

\subsubsection{Crystal free energies}
As we will see, there are two kinds of crystal lattices formed in these systems: i) simple cubic crystal lattices with ``extended'' vacancies, as found in Ref. \cite{Smallenburg2012} and ii) all other crystal phases whose vacancy concentrations are expected to be low as they only occur at high densities.  We follow slightly different routes to determine the free energy of these two types of crystals, which we describe in the following.

{\it Simple cubic lattices:} In  Ref. \cite{Smallenburg2012}, we determined the free energy of the simple cubic lattices of hard cubes, corresponding to a slant angle of $90^\circ$, for a wide range of densities and vacancy concentrations.  In order to take advantage of these ``known'' free energies for the simple cubic lattice with vacancies, we perform, similar to Ref. \cite{van2015digital}, thermodynamic integration over the shape of the particle at constant $N$, $V$, and $T$:
\begin{equation}
F(\theta_2) = F(\theta_1)+\int^{\theta_2}_{\theta_1}\left(\frac{\partial F}{\partial \theta}\right)_{NVT} d\theta. \label{eq:shape}
\end{equation}
More details on how this integration is performed are given in Section \ref{sec:shapeint}.  The integration over the particle shape yields reference free energies $F_\mathrm{ref}$ for different vacancy concentrations at a specific density $\rho_\mathrm{ref}$. We combine these with standard thermodynamic integration over the equation of state at fixed vacancy concentration and particle shape
\begin{equation}
\dfrac{\beta F(\rho)}{N}=\dfrac{\beta F_\mathrm{ref}(\rho_\mathrm{ref})}{N}+\beta \int^{\rho }_{\rho_\mathrm{ref}} d\rho' \dfrac{p(\rho')}{\rho'^ 2},
\label{eq:thermodynamicint}
\end{equation}
in order to determine the free energy as a function of shape, vacancy concentration and density.

{\it All other lattices:} For all other crystal lattices, we calculate reference free energies using the Einstein molecule method (EM), as described in Refs. \cite{vega2007revisiting,vega2008determination}. We then perform standard thermodynamic integration (Eq. \ref{eq:thermodynamicint}) using the equations of state in order to determine the free energy as a function of density for all candidate crystal phases. 

Note that for the crystal phases, we use MC simulations in the isotension-isothermal (anisotropic $NpT$) ensemble. This ensemble, in which the simulation volume can deform, is required for accurate simulation of crystal systems for which the lattice parameters are not known \textit{a priori} \cite{parrinello1981polymorphic,martovnak2003predicting,filion2009efficient}.

After determining the free energies for all competing phases, we determine phase coexistences via common-tangent constructions, and use these to draw the equilibrium phase diagram.

\subsection{Integration over particle shape}
\label{sec:shapeint}
In order to perform the integration in Eq. \ref{eq:shape}, we need to determine the derivative $(\partial F / \partial \theta)$. To do this, we use a finite difference scheme evaluating the free-energy difference between two values of $\theta$. Note that this method is reminiscent of the lattice-switch method of Ref. \cite{bruce1997free}. For simplicity, we use a simple central-difference scheme:
\begin{equation}
\frac{\partial F}{\partial \theta}\approx\frac{F({\theta+\Delta\theta/2})-F({\theta-\Delta\theta/2})}{\Delta\theta}.\label{eq:findif}
\end{equation}
As the free energy is related to the volume of phase space accessible to certain macrostates of a system, a free-energy difference between two macrostates can be expressed in terms of the ratio of the partition functions of these macrostates and, consequently, as a ratio of probabilities. Specifically, consider two macrostates $a$ and $b$, where in macrostate $a$ all particles have a slant angle $\theta_a$ and in macrostate $b$ all particles have  $\theta_b$. For fixed $N$, $V$ and $T$ the free-energy difference between these two states is then given by the ratio of the corresponding canonical partition functions $Z$:
\begin{equation}
F({\theta_a})-F(\theta_b)= k_B T \ln \left[ \frac{Z(\theta_b)}{Z(\theta_a)}\right].
\label{eq:FreeEnergyZRatio}
\end{equation}

In order to sample this ratio, we construct a simulation in which the system is allowed to switch between different slant angles. Specifically, we  introduce a MC move which switches the slant angle of all particles between $\theta_a$ and $\theta_b$. In other words, we perform the simulation in an expanded ensemble in which the system can sample two different values of $\theta$. The ratio of the two partition functions in Eq. \ref{eq:FreeEnergyZRatio} can now simply be related to the probability $P(\theta)$ of observing each slant angle in the simulation \cite{bruce1997free}: 
\begin{equation}
\frac{Z(\theta_b)}{Z(\theta_a)} = \frac{P(\theta_b)}{P(\theta_a)}.
\label{eq:FreeEnergyProbRatio}
\end{equation}

In general, the shape-switch MC move is unlikely to be accepted for anything but the smallest changes in shape. Specifically, changing the shape of all particles simultaneously will most likely create at least one overlap.   This is a similar problem to the one encountered for the lattice-switch method from Ref. \cite{bruce1997free}. One option for improving this is to introduce biasing schemes, such as umbrella sampling \cite{torrie1977nonphysical} or the multicanonical method \cite{Berg1992}. However, in our case the step size $\Delta\theta$ is a continuous variable so we can simply choose a smaller step size for which we find a reasonable acceptance rate. Choosing the step size to be small has the added advantage of improving the accuracy of our finite difference scheme. 

\begin{figure}
\includegraphics[width=.99\columnwidth]{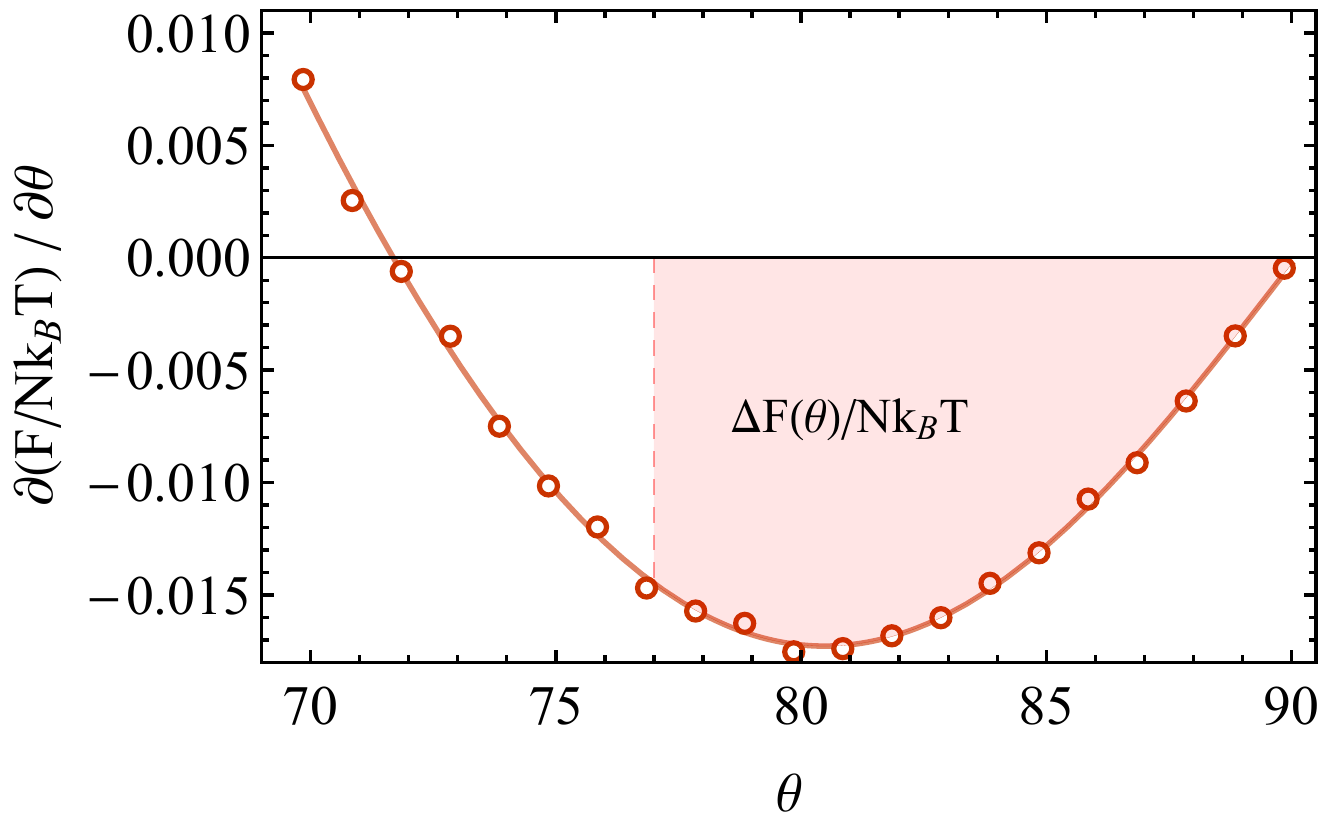} 
\caption{\label{fig:Conjugatemufit} Typical example of the behavior of the derivative $\partial F / \partial \theta$ (in dimensionless units of $k_B T$ per particle) in a defect-free simple cubic crystal of slanted cubes at a density $\rho \sigma^3=0.58$. The solid line is a polynomial fit through the data points, and the shaded area corresponds to the free-energy difference $\Delta F(\theta)= F(90^\circ)-F(\theta)$.}
\vspace{-4pt}
\end{figure}

By performing shape-switch simulations for a range of different slant angles $\theta$ and integrating over the results, we can probe the free-energy difference as a function of the slant angle $\theta$. As an example, we plot in Fig. \ref{fig:Conjugatemufit} the behavior of $\partial F / \partial \theta$ for a typical choice of density and slant angle. Integrating over this curve from one angle to another then gives the total free-energy difference between the two. To verify that this algorithm yields the same result as existing methods (i.e. the Einstein molecule method), we perform three sets of calculations: an EM calculation for a simple cubic crystal of cubes ($\theta=90^\circ$), an EM calculation for a simple cubic crystal of slanted cubes ($\theta=78.4630^\circ$), and a shape-switch (SS) calculation going from the former to the latter angle. We  perform all three calculations at the same constant number density $\rho\sigma^3=0.56$. The results of this comparison are shown in Table \ref{tab:ShapeSwitchComparison}, and one can see that the two methods agree within $0.02\ k_B T$ per particle, which is within our error bars.

\begin{table}[t]
\caption{\label{tab:ShapeSwitchComparison} Free-energy difference between a simple cubic crystal with $\theta=90^\circ$ and $\theta=78.4630^\circ$ at a density of $\rho\sigma^3=0.56$ without vacancies, as calculated by the Einstein molecule (EM) and shape-switch (SS) methods.}
\setlength{\tabcolsep}{1.6em} 
	{\renewcommand{\arraystretch}{1.3} 
		\begin{tabular}{| c | c |}
		\hline
		&  $\beta F/N$\\
		\hline
		$F_{78^\circ}$ (EM) & 7.574(15)\\
		$F_{90^\circ}$ (EM) & 7.492(15)\\
		$\Delta F_{EM} = F_{90^\circ} - F_{78^\circ}$ & -0.081(21)\\
		\hline
		$\Delta F_{SS} = F_{90^\circ} - F_{78^\circ}$ & -0.100(5)\\
		\hline
		\end{tabular}
	}
\end{table}

\section{Results}
\label{sec:CrystalPhases}

In the following we predict the phase behaviour of hard ``slanted'' cubes. Specifically, we determine candidate (crystalline) phases,  examine the associated equations of state for the system, study the vacancies in these systems, and draw the phase diagram in the packing fraction-slant angle representation.

\subsection{Candidate crystal phases}

\begin{figure}
\begin{tabular}{p{0.33\columnwidth} p{0.33\columnwidth}p{0.33\columnwidth}}
		\begin{center}{SC}\end{center} & \begin{center}{RC}\end{center} & \begin{center}{Zig-zag}\end{center}\\
		\includegraphics[width=0.33\columnwidth]{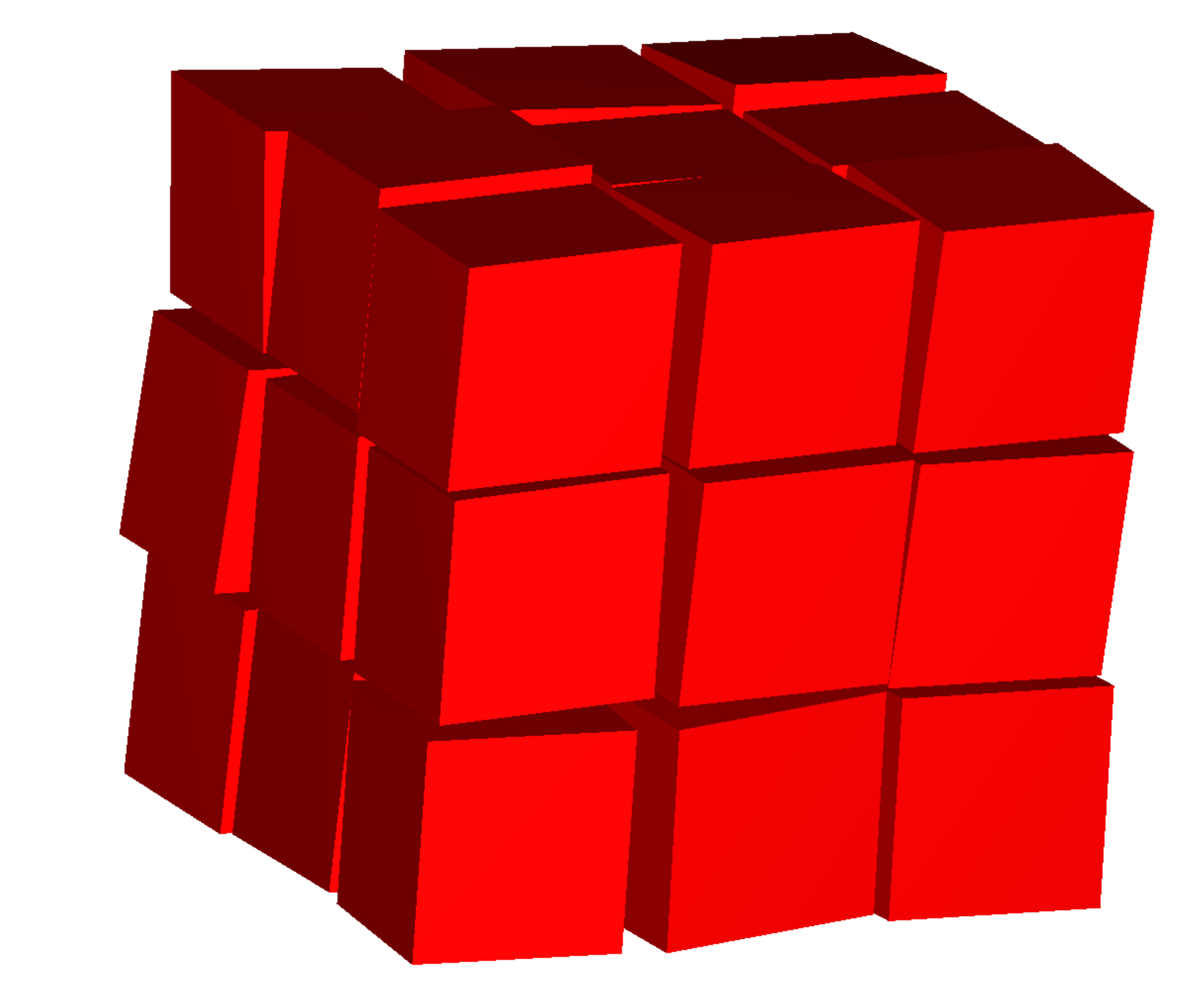}&
		\includegraphics[width=0.33\columnwidth]{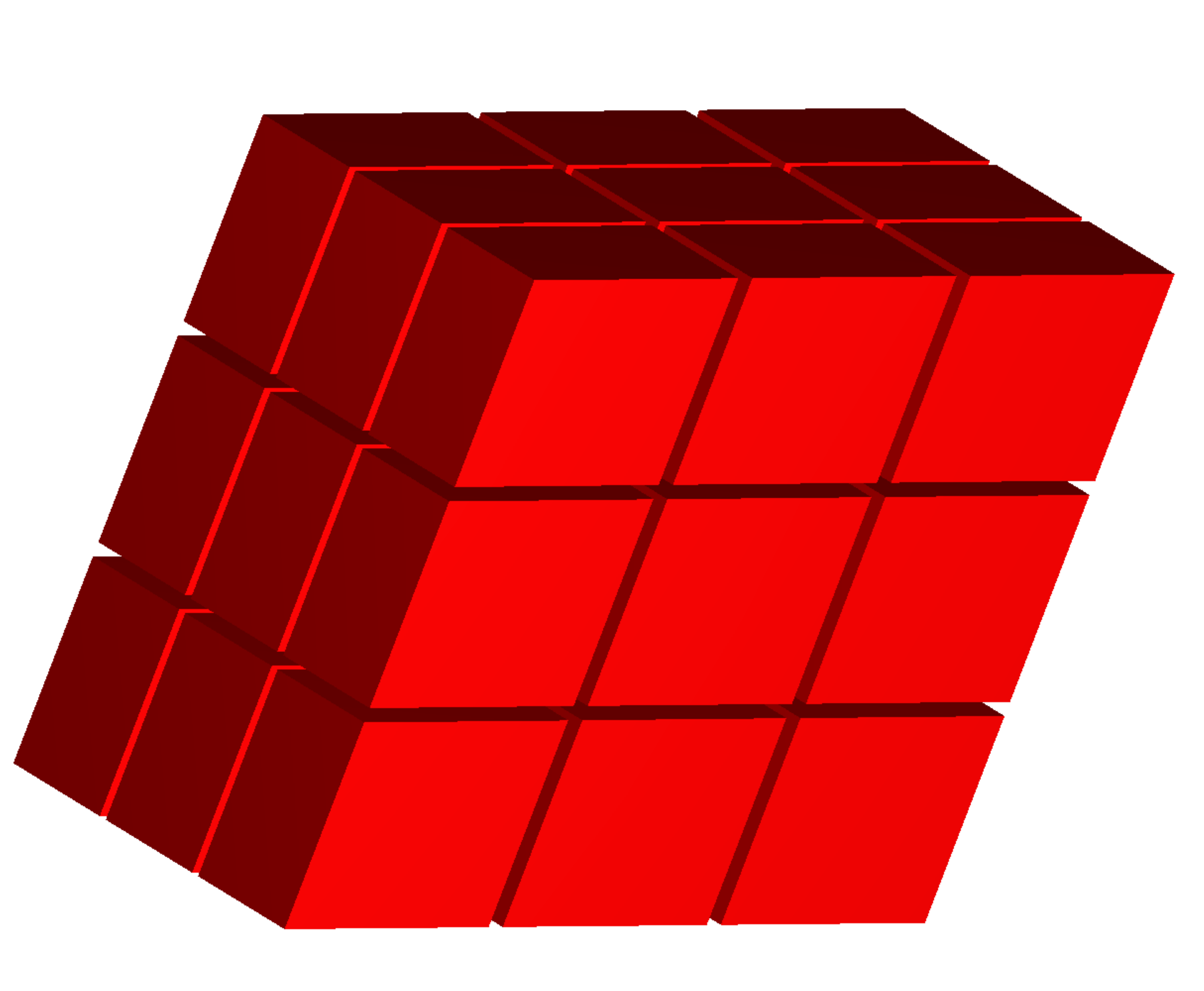}&
		\includegraphics[width=0.33\columnwidth]{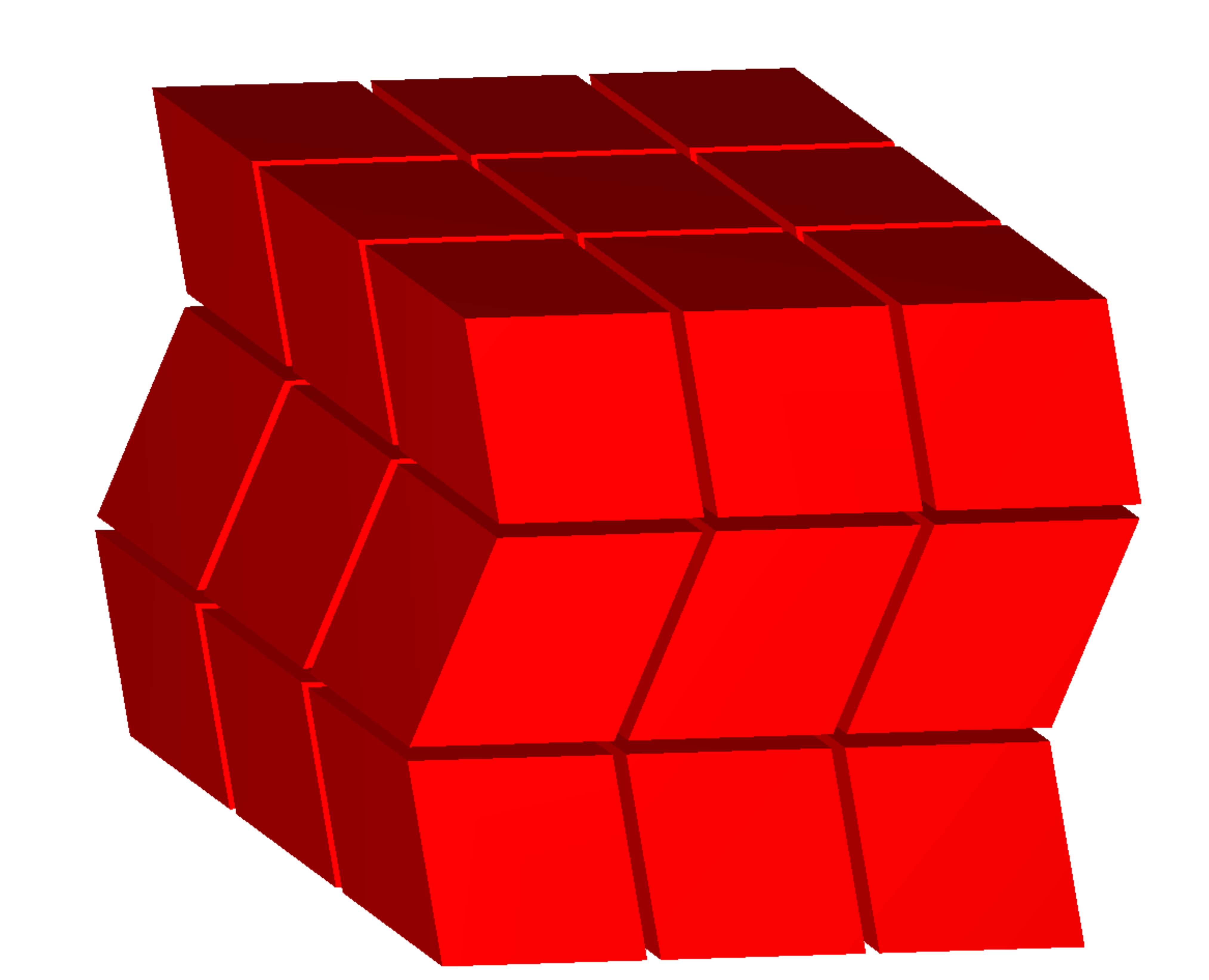}
\end{tabular}
\captionof{figure}{\label{fig:Crystals} Likely candidate crystal structures for the slanted cube particle shape. From left to right: a plastic simple cubic crystal, a rhombic crystal and a zig-zag crystal. }  
\end{figure}

In order to draw the phase diagram for this system, we first have to determine which crystalline phases are likely to be stable, i.e. we need to determine ``candidate'' crystal phases for our free-energy calculations. Due to the shape of the particles we postulate that there are three likely candidate crystal structures, namely a plastic simple cubic crystal, a rhombic crystal  and a zig-zag crystal, as illustrated in Figure \ref{fig:Crystals}. Note that both non-plastic crystals are space filling. To explore the (meta)stability of these phases, we performed MC simulations in the isotension-isothermal ensemble. Such simulations allow the box shape to transform, potentially facilitating changes in the crystal lattice. 
Interestingly, for all slant angles between $66^\circ$ and $90^\circ$,  independent of which initial crystal we chose, the system transformed into a plastic simple cubic crystal for intermediate densities. At high densities, the crystal phases always maintained the initial structure, and at low densities it melted into a fluid. The equations of state for the rhombic crystal and the zig-zag crystal were essentially indistinguishable. Further free-energy calculations revealed that the rhombic crystal was stable over the zig-zag crystal both near the coexistence region and at higher densities (see Table \ref{tab:FreeEnergiesCrystals}) with a difference of $\Delta F \approx 0.01\ k_B T /N$. For reference, the free-energy difference between the hard sphere hcp and fcc phases near coexistence is an order of magnitude smaller, $\Delta F_{hcp,fcc}\approx 0.001\ k_B T / N$ \cite{bruce1997free}. Given these results, we will only consider the rhombic crystal as the high-density phase in the following. Note that no other crystal structures were ever observed to form in our simulations.

\begin{table}[b]
\caption{\label{tab:FreeEnergiesCrystals} Reduced free energies of the rhombic and zig-zag crystal structures for packing fractions $\eta=0.75$ and $\eta=0.90$. For all crystals listed here $N=1000$.}
\setlength{\tabcolsep}{1.6em} 
	{\renewcommand{\arraystretch}{1.3} 
		\begin{tabular}{| c | c | c |}
		\hline
		& $\eta$ & $\beta F/N$\\
		\hline
		Rhombic $(72.5424^\circ)$ & 0.90 & 22.011(11)\\
		Zig-zag $(72.5424^\circ)$ &  & 22.035(11)\\
		$\beta (F_{RC} - F_{ZZ})/N$ & & -0.024(16)\\
		\hline
		Rhombic $(72.5424^\circ)$ & 0.75 & 14.228(7)\\
		Zig-zag $(72.5424^\circ)$ &  & 14.237(7)\\
		$\beta (F_{RC} - F_{ZZ})/N$ & & -0.011(10)\\
		\hline
		\end{tabular}
	}
\end{table}

To further explore the phase transitions in this system, we determine the full equation of state for each of the postulated crystal phases, as well as the fluid. We do this using a combination of event-driven molecular dynamics simulations in the $NVT$ ensemble and Monte Carlo simulations in the $NpT$ ensemble. The results are summarized in Fig. \ref{fig:EoS}. In all cases, we clearly observe a first-order phase transition from the fluid to the plastic simple cubic crystal. Note that very little hysteresis is observed in this phase transition, indicating a fairly low surface tension.  The equations of state at high densities also hint at a second first-order phase transition from the plastic simple cubic phase to a rhombic crystal. However, here we only observed a rhombic crystal melting into a simple cubic crystal, but not the reverse process. Nonetheless, further free-energy calculations confirm the presence of a second first-order phase transition as indicated in Fig. \ref{fig:EoS}.

In order to further explore the plastic nature of the simple cubic lattice, in Fig. \ref{fig:MainPhases} we show snapshots of the two crystals for a slant angle $\theta=72.5424^\circ$.  Upon visualizing the particle orientations on the unit sphere we can see that the simple cubic crystal is plastic in nature (Fig. \ref{fig:MainPhases}, bottom). Clearly there is not one preferred orientation of the particle in the simple cubic crystal. Instead, particles align their cardinal axes ($\hat{\mathbf{x}}$, $\hat{\mathbf{y}}$ and $\hat{\mathbf{z}}$ in Fig. \ref{fig:TiltedCubeVectors}) along the lattice directions, and rotate randomly between the different discrete orientations that satisfy this alignment. Note that the simple cubic crystals of (non-slanted) cubes are also plastic in nature, but due to the symmetry of cubes, these discrete orientations are identical. In contrast, in the rhombic crystal phase, the particles are all aligned perfectly and lack the freedom to rotate from one discrete orientation to the other.

\begin{table}[h!]
\begin{tabular}{p{0.48\columnwidth} p{0.48\columnwidth}}
		
		\includegraphics[width=0.48\columnwidth]{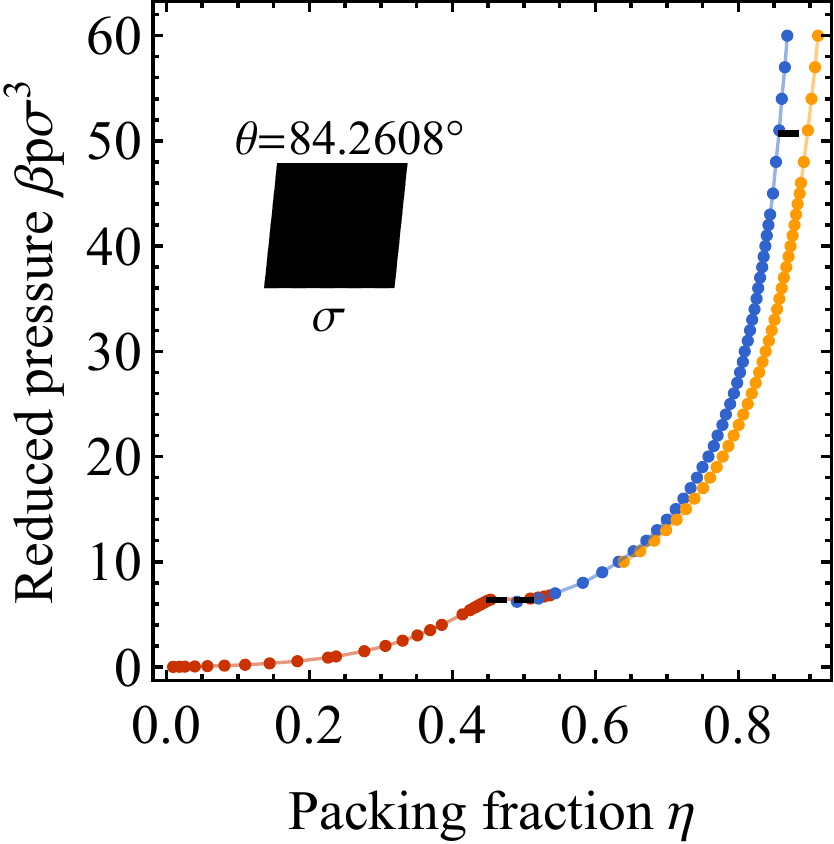}&
		\includegraphics[width=0.48\columnwidth]{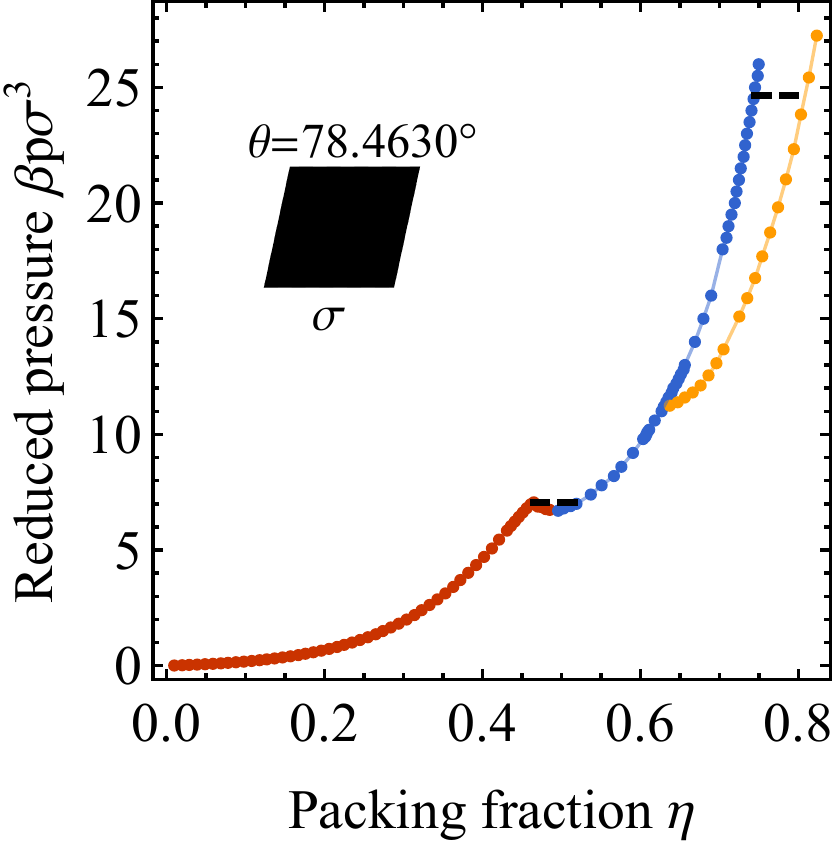}\\
		
		\includegraphics[width=0.48\columnwidth]{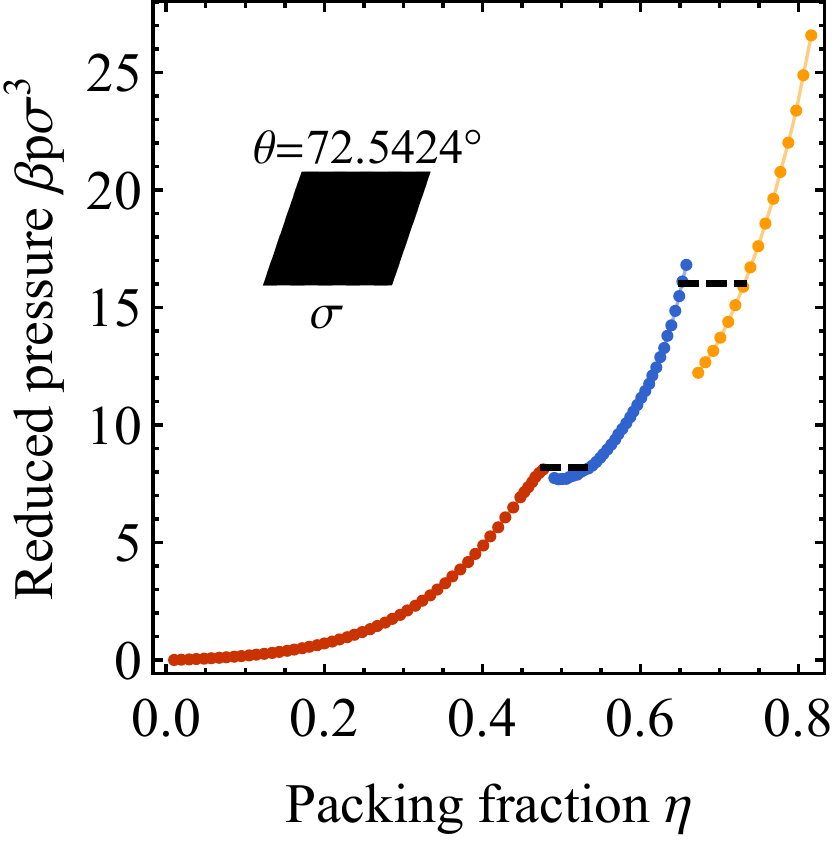}&
		\includegraphics[width=0.48\columnwidth]{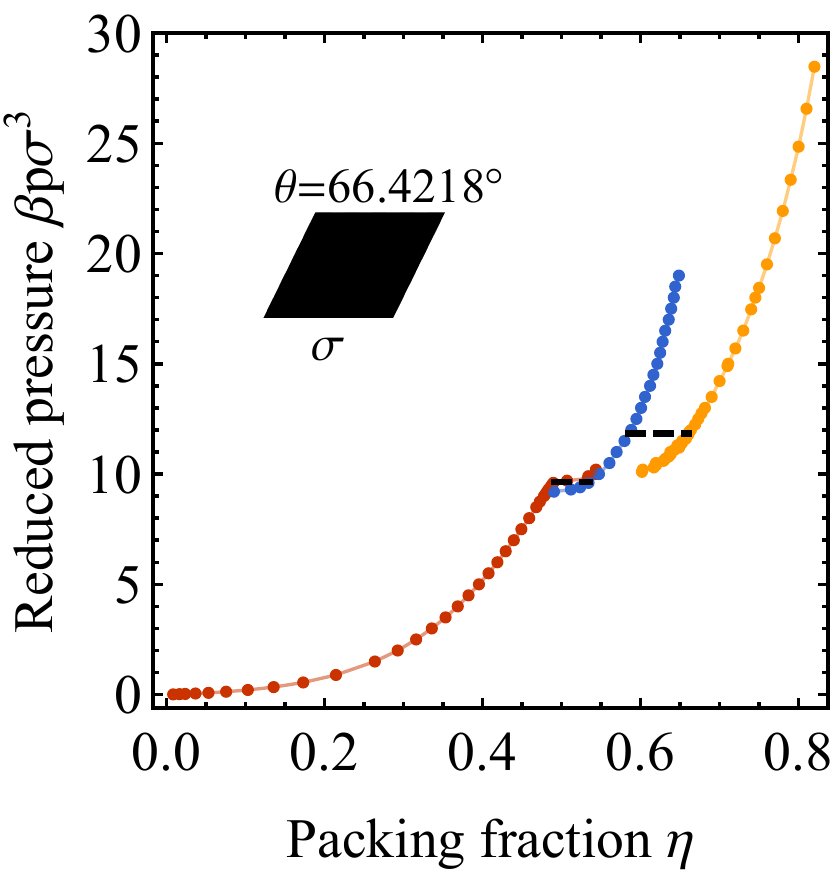}\\
		 
\end{tabular}
\captionof{figure}{\label{fig:EoS} Equations of state for various slant angles. For all slant angles we find three phases. Namely, at low densities a fluid (red), at intermediate densities a simple cubic crystal phase (blue), and at high densities a rhombic crystal phase (yellow). Dashed black lines indicate the coexistence densities and pressure as determined by free-energy calculations.}  
\end{table}

\begin{table}[h!]
\begin{tabular}{p{0.48\columnwidth} p{0.48\columnwidth}}
		\begin{center}{\large Simple Cubic (SC)}\end{center} & \begin{center}{\large Rhombic Crystal (RC)}\end{center}\\
		\includegraphics[width=0.48\columnwidth]{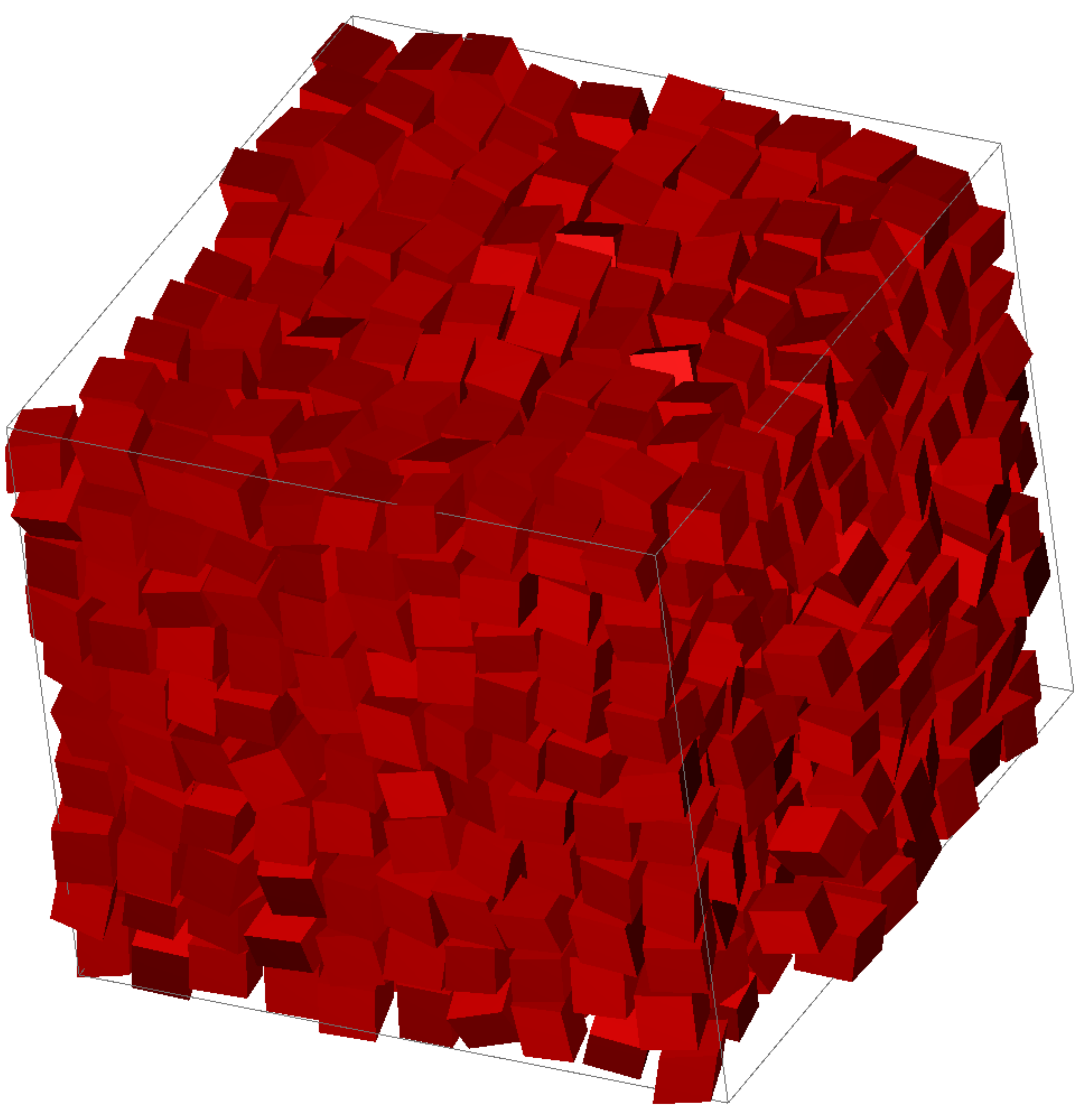}&
		\includegraphics[width=0.47\columnwidth]{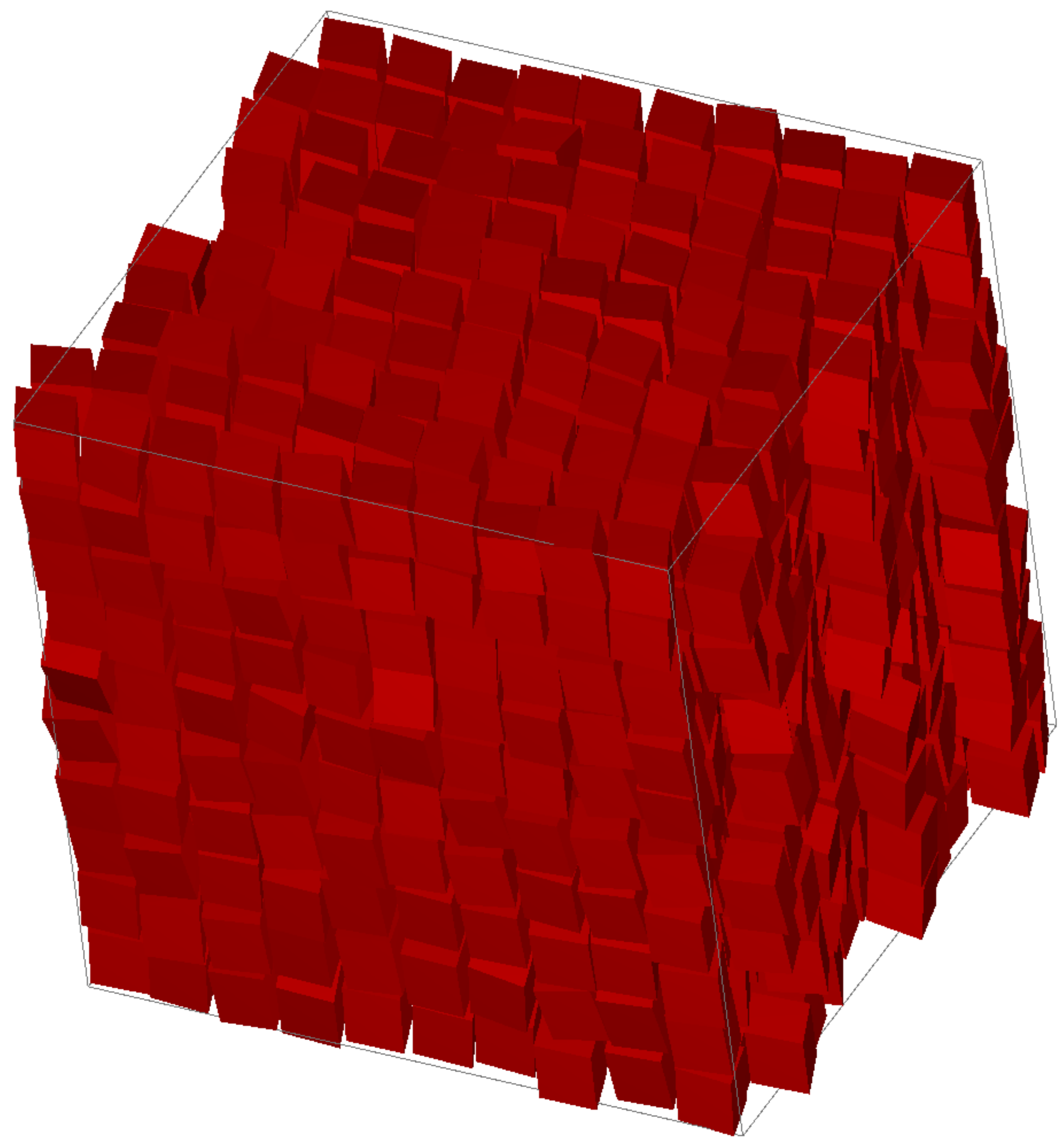}\\
		
		\includegraphics[width=0.47\columnwidth]{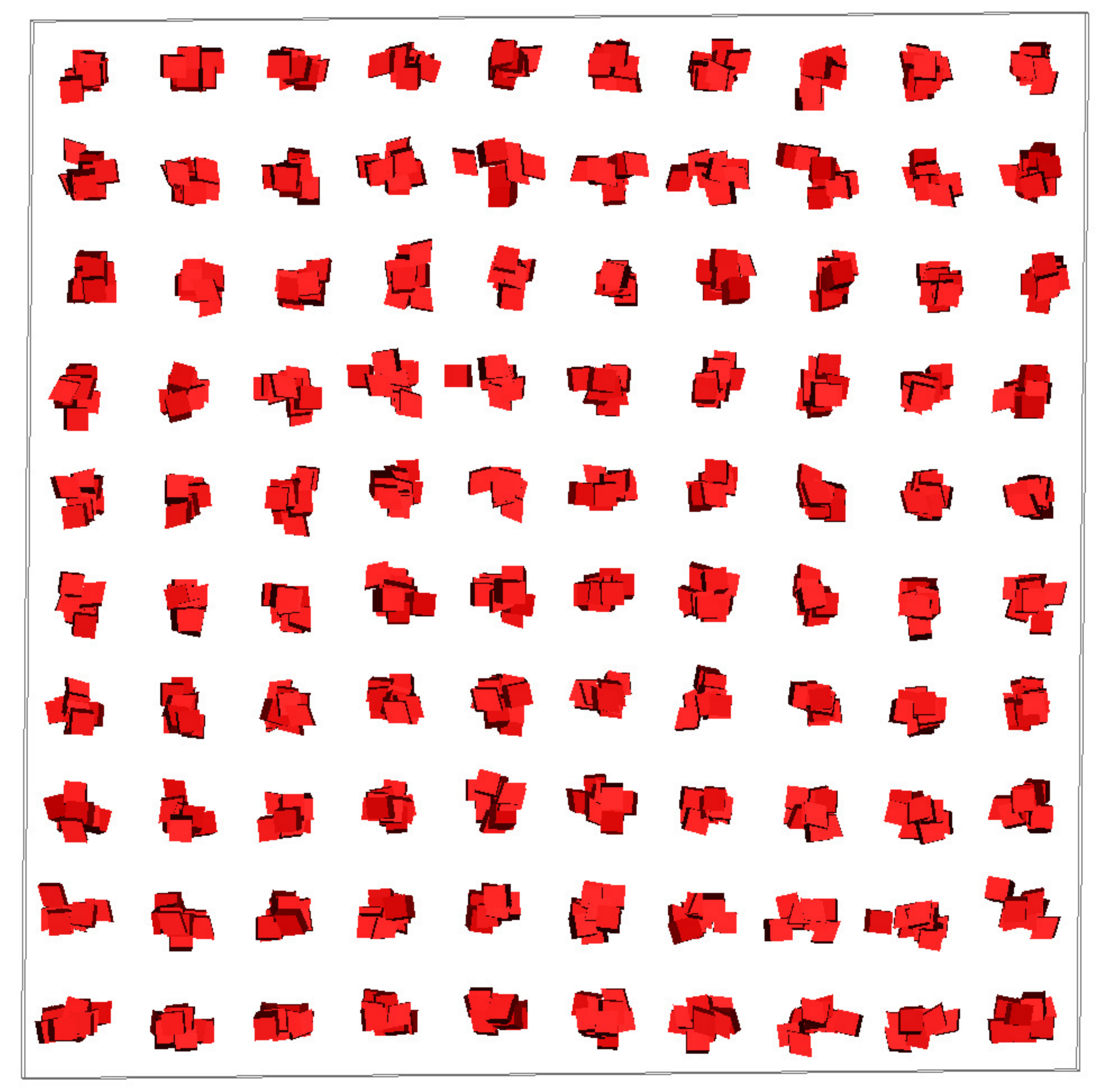}&
		\includegraphics[width=0.485\columnwidth]{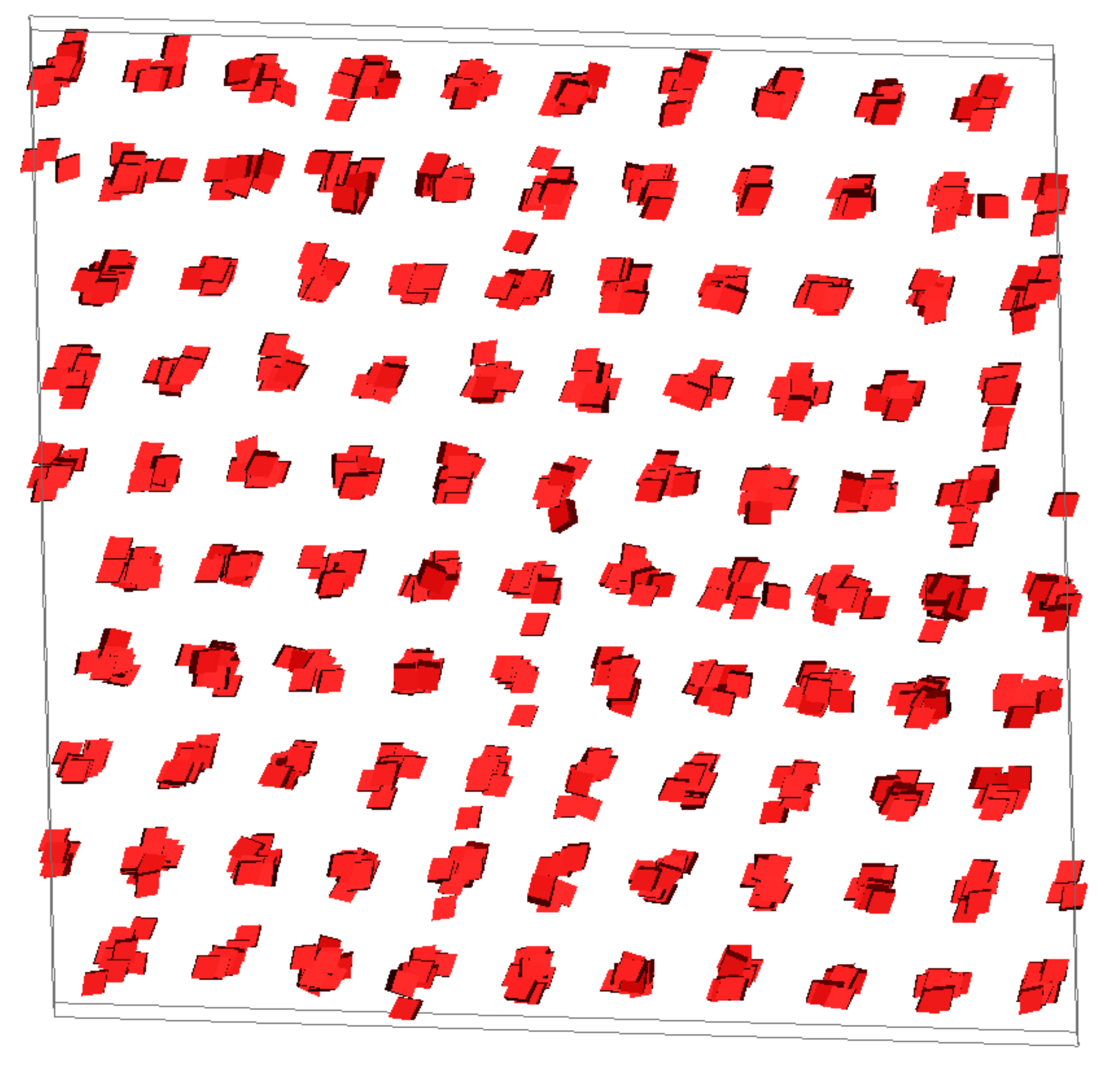}\\
		
		\includegraphics[width=0.48\columnwidth]{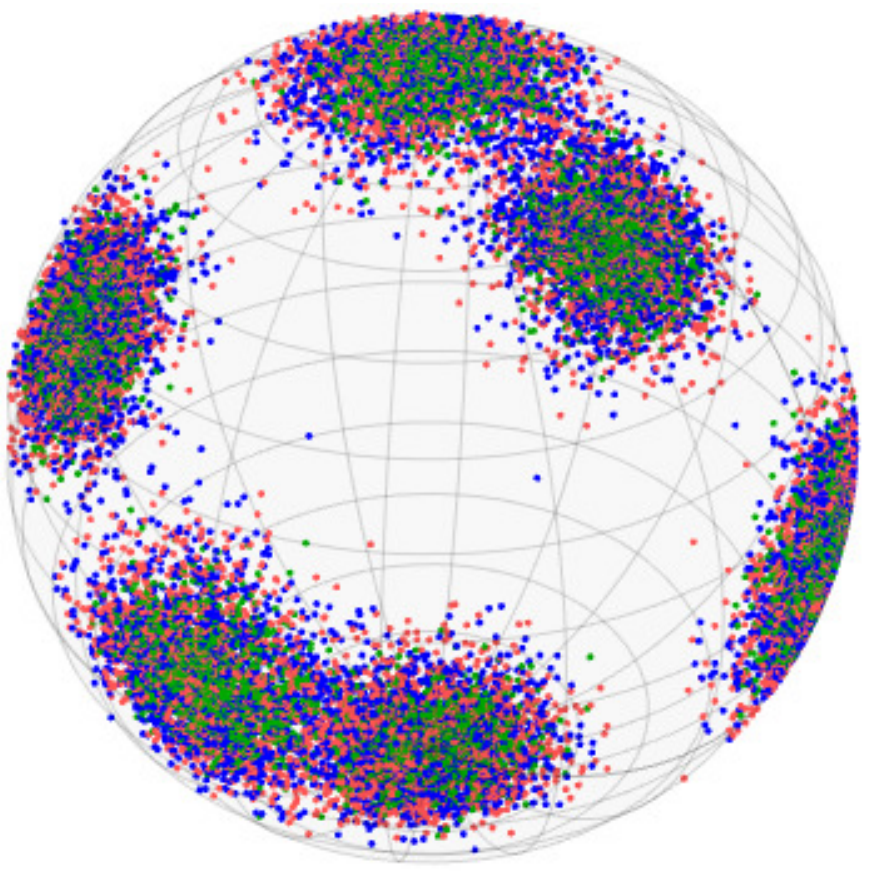}&
		\includegraphics[width=0.48\columnwidth]{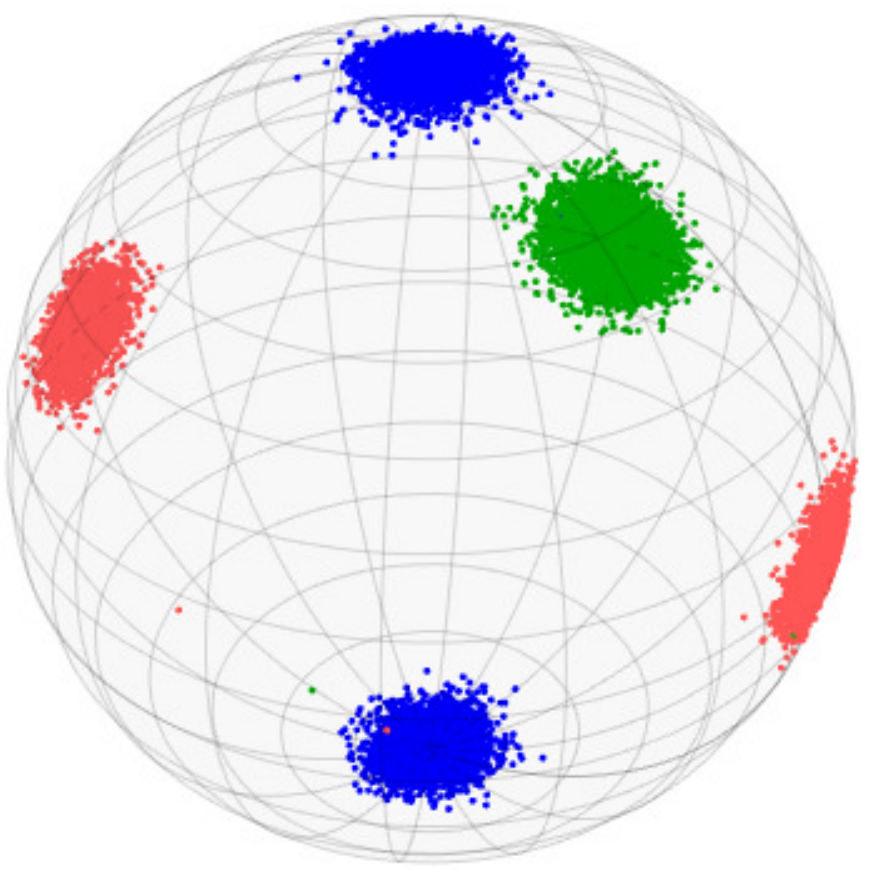}
\end{tabular}
\captionof{figure}{\label{fig:MainPhases} Crystal phases of slanted cubes. Top row: snapshots with particles  depicted at their true size. Middle row: particles shrunk to a fraction of their true size to show the crystal lattice. Bottom: Scatter plot of the particle orientations. Red, green and blue indicate respectively the $\hat{\mathbf{x}}$, $\hat{\mathbf{y}}$ and $\hat{\mathbf{z}}$ unit vectors in the reference frame of the particle (Fig. \ref{fig:TiltedCubeVectors}).
The left column shows the simple cubic crystal phase and the right column the rhombic crystal phase.}
\end{table}
 
\newcommand{\vac}{\mathrm{vac}}
\newcommand{\comb}{\mathrm{comb}}
\subsection{Vacancy concentration}
\label{sec:Vacancies}
We now turn our attention to the vacancies in the simple cubic crystal. As the simple cubic lattice of hard cubes has an extremely high equilibrium vacancy concentration near coexistence, we postulate that vacancies will also be very important for the plastic simple cubic crystal formed by our ``slanted'' cubes. 

\begin{figure}[h!]
\includegraphics[width=.99\columnwidth]{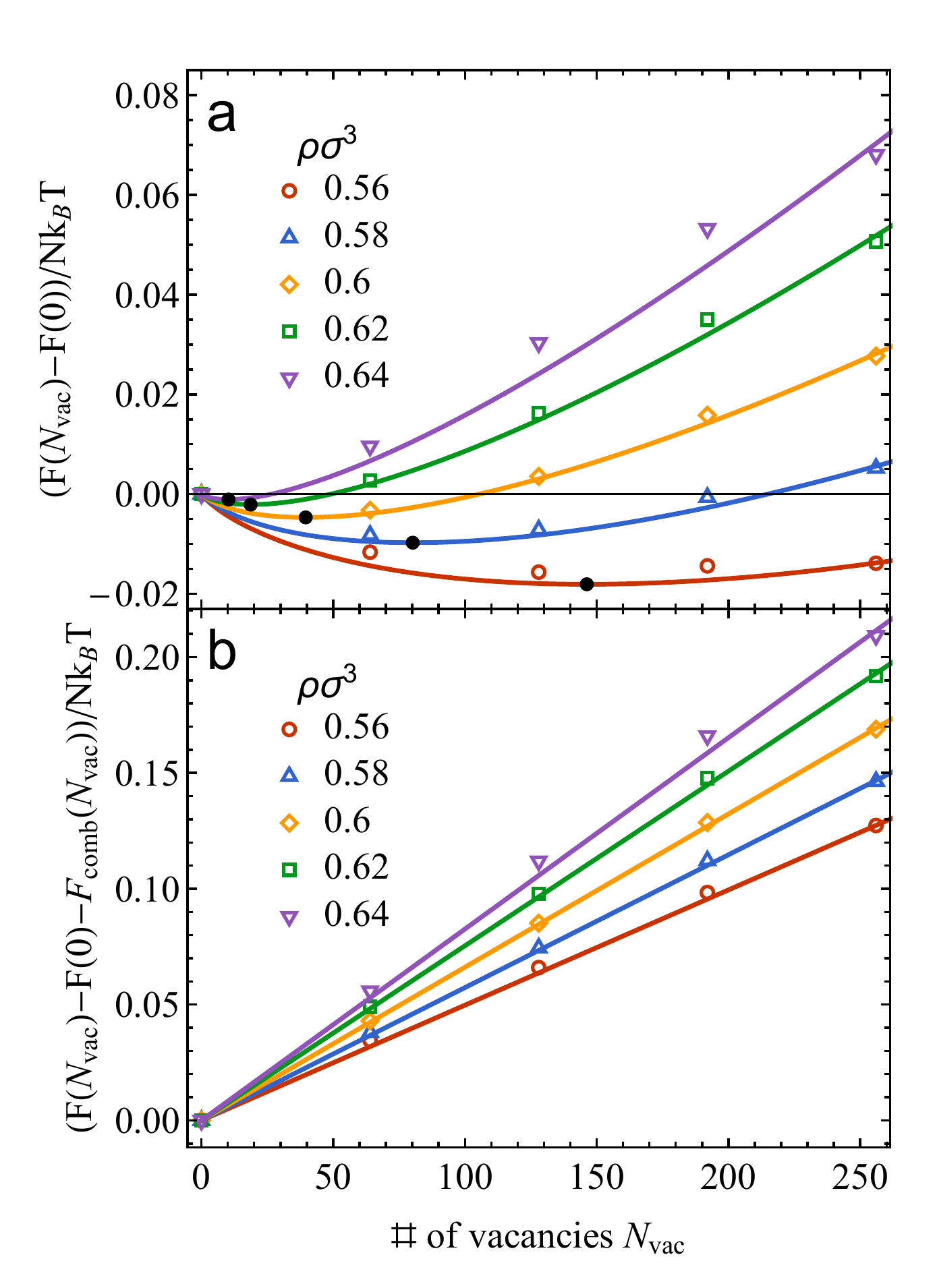}\\
\caption{\label{fig:LinearDefectF} (a) Fits through the vacancy contributions to the free energies of an SC crystal of slanted cubes with slant angle $\theta=78.463^\circ$ with $N_L = 8000$ lattice sites. The fitted value for $F(0)$ has been subtracted to plot all curves in the same graph. Open symbols indicate the data, solid lines the fits, and filled (black) symbols the minimum of these fits: the equilibrium vacancy concentration. (b) The same as top, but with the combinatorial term subtracted as well. The linear dependence on the number of vacancies can be seen clearly here.}
\end{figure}

To this end, we calculate equations of state and free energies for the simple cubic crystal for slant angles $\theta=78.4630^\circ$ and $\theta=72.5424^\circ$ and vacancy concentrations $\alpha=0.000$, $0.008$, $0.016$, $0.024$ and $0.032$. Here, $\alpha$ is defined as the fraction of lattice sites not filled by a particle, i.e. $\alpha = N_\vac /N_L$, with $N_\vac$ the number of vacancies and $N_L$ the number of lattice sites. The shape-switch calculations were also performed at these vacancy concentrations, at a constant density of $\rho\sigma^3=0.58$, as at this density the simple cubic crystal is stable for all slant angles $66^\circ\leq\theta\leq 90^\circ$. By subtracting the free energy of a crystal lattice without vacancies $F_\mathrm{perf}$ from the total free energy $F$, we can look purely at the contribution of the vacancies. This is shown in Fig. \ref{fig:LinearDefectF}a, where we plot the free energy $F(N_\mathrm{vac})$ as a function of the number of vacancies $N_{\mathrm{vac}}$, for a slant angle $\theta=78.463^\circ$. We can clearly see that for slanted cubes, the free energy of the simple cubic crystal phase is lowered by incorporating a substantial fraction of vacancies. We thus find, similar to the hard cubes \cite{Smallenburg2012} and truncated hard cubes \cite{gantapara2013phase}, a vacancy-rich simple cubic crystal.

Typically, the effect of vacancies on the free energy of a crystal can be decomposed into a free-energy cost for creating the vacancy, and a combinatorial entropy associated with the location of the empty lattice site. When vacancies in a crystal do not strongly interact (as was shown to be the case for hard cubes \cite{Smallenburg2012} and truncated hard cubes \cite{gantapara2013phase}), this can be written as
\begin{equation}
 F(N_\vac) = F(0) + \epsilon_\vac N_\vac + F_\comb (N_\vac)
\end{equation}
where $\epsilon_\vac$ is the free-energy cost of creating a defect, and the second term is the combinatorial entropy
\begin{equation}
 F_\comb (N_\vac)= -k_B T \log \left[\frac{N_L!}{N!(N_L-N)!}\right],
\end{equation}
with $N_L$ the number of lattice sites. 
If we subtract the combinatorial term from the data in Fig. \ref{fig:LinearDefectF}a, we indeed clearly see a linear dependence of $F(N_\vac)$ on $N_\vac$, indicating that the vacancies also only interact weakly in crystals of slanted cubes (see Fig. \ref{fig:LinearDefectF}b). 

\begin{figure}[h!]
\includegraphics[width=0.9\columnwidth]{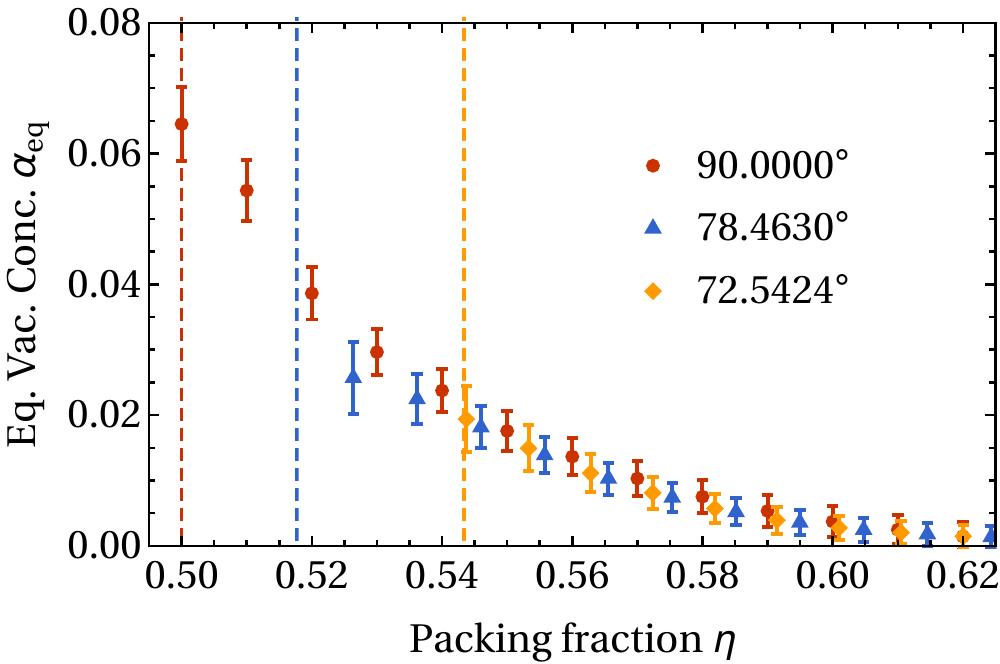}
\caption{\label{fig:eqvacpl} The equilibrium vacancy concentration $\alpha_{eq}$ as a function of the packing fraction $\eta$. The dashed lines indicate the melting densities. The equilibrium vacancy concentration is independent of the slant angle when compared at the same packing fraction.}
\end{figure}

From fits to $F(N_{\mathrm{vac}})$, such as those shown in Fig. \ref{fig:LinearDefectF}a, we obtain the equilibrium vacancy concentration at each density, i.e. the vacancy concentration $\alpha$ which minimizes the free energy. We summarize these results as a function of the slant angle and the packing fraction $\eta$ in Fig. \ref{fig:eqvacpl}. Amazingly, we can see that the equilibrium vacancy concentration of the simple cubic lattice is essentially independent of the slant angle $\theta$. We note that while such an invariance to shape was also found for vacancies in a simple cubic crystal of parallel rounded cubes \cite{marechal2011phase}, it was not found for a simple cubic crystal of cuboctahedra \cite{gantapara2013phase}.

\subsection{Phase diagram}
\label{sec:PhaseDiagram}

\begin{figure}[t]
\includegraphics[height=0.2\textheight]{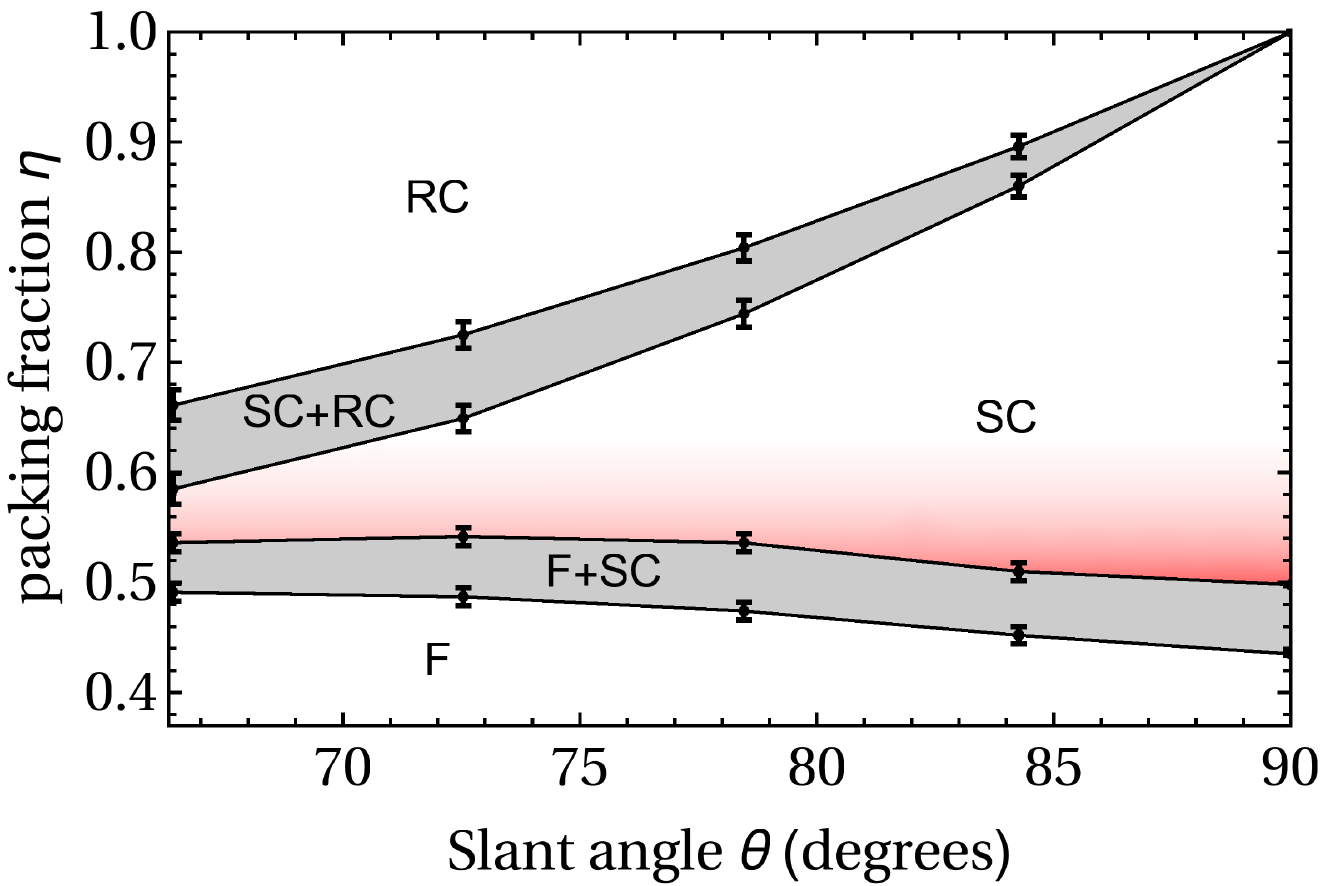}
\includegraphics[height=0.2\textheight]{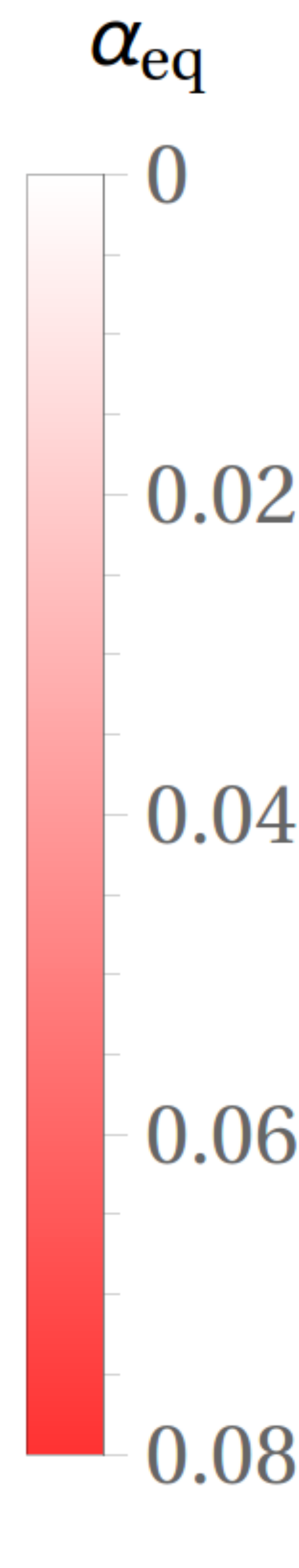}
\caption{\label{fig:PhaseDiagram} Phase diagram in the $\theta$-$\eta$ plane for the slanted cube system. Indicated are the fluid (F), simple cubic crystal (SC) and rhombic crystal (RC) phases. The gray shaded areas denote coexistence regions and the (red) colouring represents the equilibrium vacancy concentration in the SC phase. The lines connect the points and merely serve to guide the eye. 
}
\end{figure}

Combining all our free-energy calculations for different slant angles, densities, and vacancy concentrations, we determine the phase diagram for this system.  Our results are shown in Fig. \ref{fig:PhaseDiagram}, where we plot the predicted phase diagram in the $\theta$-$\eta$ plane. For the sake of completeness we also plot the equilibrium vacancy concentration of the simple cubic phase through the use of a colour map. As expected from our observations of the equations of state, we observe a stable simple cubic phase over the entire investigated range of slant angles, which coexists with a fluid at low packing fraction, and with a rhombic crystal at high packing fraction. The stability range of the simple cubic phase decreases with decreasing slant angle, which can be intuitively understood as the increasing asymmetry of the particles becomes more and more incommensurate with simple cubic ordering. As $\theta$ decreases, the particles become less and less cube-like, and as a result neighboring particles in conflicting orientations increasingly interfere with each others freedom of movement in the crystal, in particular at high densities. The resulting decrease in entropy destabilizes the simple cubic phase with respect to both the fluid and the rhombic crystal phase, leading to a smaller area of stability.

\section{Conclusions}
\label{sec:Conclusions}
In conclusion, we examined the phase behaviour of a system of hard slanted cubes, and in particular focused on the role of vacancies in the equilbrium phase behaviour for this system. We find three stable phases for slant angles $66^\circ \leq \theta \leq 90^\circ$, namely a fluid 
phase, a vacancy-rich simple cubic crystal phase, and a rhombic crystal phase. Note that we find that the vacancy-rich simple cubic phase always appears at intermediate densities, while the rhombic phase occurs at high densities. 

Interestingly, we find that for the rhombic crystal the lattice angles are strictly identical to the particle slant angle. This is in sharp contrast to two-dimensional self-assembly experiments of rhombic platelets~\cite{zhao2012twinning}, where it was shown that the angle of the rhombic lattice can differ significantly from the slant angle of its constituent particles. We hypothesise that this disparity between experiment and simulation may be attributed both to rounding of particle edges and depletion interactions in experiments.  It is well known that particles with rounded edges and/or corners can form crystal structures in which the lattice angles differ significantly from the perfectly sharp shape. Rounded squares, cubes and superballs are the most notable and well-studied examples of this, forming lattices with different angles depending on their rounding \cite{rossi2015shape,avendano2012phase,zhao2011entropic}.

The more surprising and important part of our results concerns the behaviour of the vacancies in the simple cubic lattice. Specifically, we find that the equilibrium vacancy concentration in the simple cubic phase is essentially independent of the slant angle - meaning that it is the same as for simple hard cubes.  Moreover, in agreement with what was seen for hard cubes, we show that the vacancies in these systems also do not strongly interact.

Our results clearly indicate that extended vacancies are extremely robust with respect to distortions in the particle symmetry, and thus might be present in a much wider range of colloidal polyhedra than the particles with cubic symmetry that have previously been studied. We note, however, that the extended defects observed in this system are still restricted to the simple cubic crystal; we do not predict significant vacancy concentrations in the rhombic crystal phase, which only occurs at much higher packing fractions. Hence,  whether extended defects can be realized in other crystalline lattices remains an open question. However, we speculate that if a defect is able to extend, the free-energy cost to create it must be low, which in turn is indicative of a high equilibrium vacancy concentration for the crystal as a whole. We aim to test this hypothesis in future work.

\section*{Acknowledgements}
\label{sec:Acknowledgements}
We thank Marjolein Dijkstra and Matthieu Marechal for useful discussions and Michiel Hermes for a careful reading of the manuscript. L. F. acknowledges funding from the Dutch Sector Plan Physics and Chemistry and financial support from the Netherlands Organization for Scientific Research (NWO-VENI grant No. 680.47.432). F. S. gratefully acknowledges funding from the Alexander von Humboldt foundation.





%

\end{document}